\documentclass[aps,twocolumn,showpacs,amsmath]{revtex4-1}

\usepackage[pdftex]{graphicx}
\usepackage{newlfont}
\usepackage{amssymb}
\usepackage{amsfonts}
\usepackage{amsmath}
\usepackage{keyval}
\usepackage{graphicx}
\usepackage{bm}
\usepackage{bbm}
\usepackage{epsfig}
\usepackage{epstopdf}
\usepackage{float}

\newcommand{\ket}[1]{|{#1}\rangle}
\newcommand{\bra}[1]{\langle{#1}|}
\newcommand{\braket}[2]{\langle {#1} | {#2} \rangle}

\usepackage{hyperref}
\usepackage{color}
\definecolor{med-blue}{RGB}{25,25,112}
\hypersetup{colorlinks, linkcolor={red},citecolor={blue}, urlcolor={blue}}

\begin{document}

%\title{Quantum simulation of unitary dynamics of an XY spin chain \\ with a dipolar coupled spin system}

%\title{Quantum simulation of mirror inversion operation in an XY spin chain \\ with a dipolar coupled spin system}

%\title{Mirror inversion operation in an XY spin chain: \\ Quantum simulation with a dipolar coupled spin system}

%\title{Simulating mirror inversion in an XY spin chain using a dipolar coupled spin system}

\title{Simulation of mirror inversion of quantum states in an XY spin chain using NMR}

\author{K. Rama Koteswara Rao$^1$, T. S. Mahesh$^2$, and Anil Kumar$^1$}

\affiliation{$^1$Centre for Quantum Information and Quantum Computation,\\
 Department of Physics and NMR Research Centre, Indian Institute of Science, Bangalore 560012, India\\
 $^2$Department of Physics and NMR Research Center, Indian Institute of Science Education and Research, Pune 411008, India}

\begin{abstract}
%We report experimental quantum simulation of the mirror inversion operation in a spin chain proposed in [Phys. Rev. Lett. \textbf{93}, 230502 (2004)]. The experiment is performed with a 5-qubit dipolar coupled spin system using nuclear magnetic resonance techniques. To perform quantum simulation we make use of the recently proposed algorithmic approach to product decompose unitary operator in the Pauli operator basis along with numerical pulse optimization techniques. We further demonstrate, using mirror inversion, entangled states can be transferred from one end of the chain to the other end.

We report an experimental quantum simulation of unitary dynamics of an XY spin chain with pre-engineered couplings. Using this simulation, we demonstrate  the mirror inversion of quantum states, proposed by Albanese \textit{et al}. [Phys. Rev. Lett. \textbf{93}, 230502 (2004)]. The experiment is performed with a 5-qubit dipolar coupled spin system using nuclear magnetic resonance techniques. To perform quantum simulation we make use of the recently proposed unitary operator decomposition algorithm of Ajoy \textit{et al}. [Phys. Rev. A \textbf{85}, 030303 (2012)] along with numerical pulse optimization techniques. Further, using mirror inversion, we demonstrate that entangled states can be transferred from one end of the chain to the other end. The simulations are implemented with high experimental fidelity, which implies that these kind of simulations may be possible in larger systems.

\end{abstract}

\pacs{03.67.Mn, 82.56.-b}

\maketitle

\section{Introduction}
Quantum transport from one register to another is one of the fundamental tasks in quantum information processing. Recently spin chains with nearest-neighbour interactions were proposed to perform this task efficiently \cite{BosePRL}. The main motivation for using these spin chains as communication channels is that one could use the natural unitary evolution of the spin chain to drive quantum information from one quantum register to another with limited or no external control. Since its proposal, many interesting protocols for transferring quantum information using spin chains with different kinds of interactions have been reported \cite{ChrPRL, ChrPRA, Fitz06, BoseCP, BurgPRA, DiFPRL, KayRW, Ajoy12}. Apart from quantum state transfer, spin chains were also proposed for transferring, distributing, and generating entanglement \cite{SubraPRA, Amico, Plastina, Furman, Wang, YungPRA}. Some of these proposals have been verified experimentally by simulating the spin chains using Nuclear Magnetic Resonance (NMR) techniques \cite{Zhang2005, Zhang2007, Jones07, Zhang2009, Alvarez, RaoIJQI}. Universal quantum computation using permanently coupled spin chains were also studied \cite{Benjamin, YungPRL, KayNJP}. 

Albanese \textit{et al}. \cite{Albanese} have shown that mirror inversion of a quantum state with respect to the centre of the chain can be achieved by modulating the couplings of XY spin chains. Apart from transferring the quantum state of a single qubit, this mirror inversion operation can be used to transfer non-trivial entangled states of multiple qubits between different registers or from one part to another within a register. In this paper, our aim is to simulate the unitary evolution of an XY spin chain with pre-engineered couplings and experimentally demonstrate the mirror inversion operation proposed in reference \cite{Albanese}. Moreover we also demonstrate experimentally that entangled states can be transferred from one end of the chain to the other using mirror inversion. 

For simulating the spin chain, we follow the digital quantum simulation approach \cite{Buluta}, where the unitary evolution of a chain is divided into a circuit consisting of one- and two-qubit gates. There are many ways by which one can decompose an arbitrary unitary operator into one- and two-qubit gates \cite{Tucci, Cartan, Vart}. Recently, Ajoy \textit{et al}. \cite{Ashok} proposed an algorithm for product decomposing an arbitrary unitary operator into a chosen operator basis. Using this method, we can obtain the unitary operator decomposition directly in the Pauli operator basis. This is advantageous as in spin based quantum architectures like NMR it is easier to implement gates from Pauli operator basis, such as ZZ, compared to the gates like controlled-NOT (C-NOT).

Here, we use a combination of this algorithm and GRadient Ascent Pulse Engineering (GRAPE) algorithm \cite{GRAPE}, which is a numerical pulse optimization technique, to simulate the unitary evolution of the XY spin chain.
Specifically, 
employing the algorithm given by Ajoy \textit{et al}. \cite{Ashok}, we first provide product decompositions of unitary evolutions of 4- and 5-spin XY chains into the Pauli operator basis.
Then, in the experiments, we realize each of these decomposed unitary operators using GRAPE technique. Also, we extend the product decompositions of unitary evolutions to the $N$-spin XY chains.
When combined with the subsystems approach given by Laflamme and co-workers \cite{Mahesh12, RyanPRA}, the methods presented here are useful in realizing quantum simulations by much larger spin systems.

%Specifically, 
%employing the algorithm given by Ajoy et al \cite{Ashok}, we first provide decompositions of unitary evolutions of 4- and 5-spin XY chains with pre-engineered couplings and then generalize it to $N$-qubits. 
%In the experiments, we realize the decomposed unitary operators of 4- and 5-spin chains using GRAPE technique.
%When combined with the subsystems approach given by Laflamme and co-workers \cite{Mahesh12, RyanPRA}, the methods presented here are useful in realizing quantum simulations by much larger spin systems.

%For the experiment, we use a 5-spin system partially oriented in a liquid crystal.
The experiments are performed by using NMR techniques on a 5-spin system, partially oriented in a liquid crystal.
The spins are thus coupled by direct dipolar interactions as well as indirect scalar couplings. The dipolar coupling strengths are an order of magnitude stronger than the scalar couplings, which are frequently used to realize multiqubit gates in liquid state NMR quantum information processing. The large dipolar couplings make the multiqubit gates faster. Also, they help the GRAPE algorithm to find  multiqubit gates with less number of time steps, which reduces the computational time for numerical optimization.

The paper is arranged as follows. In sec II, we describe the mirror inversion operation in XY spin chains with pre-engineered couplings and the product decomposition of their unitary evolution into the Pauli operator basis. In sec III, we present the experimental implementation and discuss the results, and in sec IV we conclude.

\section{Mirror Inversion}
\label{sec-mirinv}

We first revisit the mirror inversion operation in spin chains proposed in references \cite{Albanese, Karbach}. Consider a chain of N spin-1/2 particles, coupled by nearest-neighbour XY interaction, with the Hamiltonian, 
\begin{equation}
\label{Hamqubit}
{\cal H}= \frac{1}{2} \sum_{i=1}^{N-1} J_i (\sigma_i^x \sigma_{i+1}^x+\sigma_i^y \sigma_{i+1}^y) + \frac{1}{2} \sum_{i=1}^N h_i (\sigma_i^z+1).
\end{equation}
Here $\sigma_i^{x/y/z}$ are the Pauli matrices for the spin $i$, $J_i$ is the coupling between the spins $i$ and $i+1$, and $h_i$ is the local magnetic field at the spin site $i$. The above Hamiltonian commutes with the total z-spin operator $\sigma_{tot}^z=\sum_{i=1}^N \sigma_i^z$, i.e., $[\sigma_{tot}^z, \cal{H}]=0$, which means that the Hilbert space of the system can be divided into invariant subspaces, characterized by distinct eigenvalues of the $\sigma_{tot}^z$ operator.

The Hamiltonian in Eq. (\ref{Hamqubit}) can be mapped to a Hamiltonian of non-interacting spinless fermions by using the following Jordon-Wigner transformation \cite{Lieb},
\begin{equation}
c_i = \left(\prod_{j<i} \sigma_j^z\right)\frac{\sigma_i^x+i\sigma_i^y}{2}; \phantom{aaa}  c_i^\dagger = \left(\prod_{j<i} \sigma_j^z\right)\frac{\sigma_i^x-i\sigma_i^y}{2},
\end{equation}
where $c_i$ and $c_i^\dagger$ are fermionic operators. 
The Hamiltonian in the second quantized form is given by
\begin{equation}
\cal{H} = \sum_{i=1}^{N-1} J_i (c_i^\dagger c_{i+1}+c_{i+1}^\dagger c_i)+\sum_{i=1}^N h_i c_i^\dagger c_i.
\end{equation}

Let $\ket{j}$ represents a single-particle state, i.e., there is a single fermion at the site $j$ and all other sites are empty. Equivalently, for spin chains, $\ket{j}$ represents a state, where the $j^{th}$ spin is in the state $1$ and all the other are in the state $0$.
In the single particle subspace i.e., the subspace spanned by $\ket{j}$, the Hamiltonian $\cal{H}$ can be represented by the following tridiagonal matrix,
\begin{equation}
\cal{H}_1=
\begin{pmatrix}
h_1 & J_1 & 0 & \cdots & 0 \\
J_1 & h_2 & J_2 & \cdots & 0 \\
0 & J_2 & h_3 & \cdots & 0 \\
\vdots & \vdots & \vdots & \ddots & J_{N-1} \\
0 & 0 & 0 & J_{N-1} & h_N
\end{pmatrix}.
\end{equation}

It is assumed that the Hamiltonian of the spin chain possess mirror symmetry, that is $J_i=J_{N-i+1}$ and $h_i=h_{N-i}$. Then, the $N$-dimensional eigenvectors of $\cal{H}_1$ will have definite parity, i.e, every eigenvector of $\cal{H}_1$ is either even or odd. Due to a well-known theorem given in reference \cite{Karbach}, the eigenvectors of $\cal{H}_1$ (a real symmetric tridiagonal matrix with only positive sub diagonal elements) are alternatively even and odd.

Perfect mirror inversion of the single-particle state implies, for some time $\tau$ and up to some global phase $\phi_0$,
\begin{equation}
\label{mirinv1}
e^{-i\cal{H}\tau} \ket{i} = e^{i\phi_0} \ket{N-i}.
\end{equation}

Let  $\ket{\nu}$ be a single particle eigenstate with eigenvalue $\varepsilon_{\nu}$. Then the single particle state $\ket{N-i}$ can be written as $\ket{N-i} = \sum_{\nu}\ket{\nu}\braket{\nu}{N-i}$. Due to the alternating parity property, $\braket{N-i}{\nu} = (-1)^{\nu} \braket{i}{\nu}$. This implies, $\ket{N-i}=\sum_{\nu}(-1)^{\nu}\ket{\nu}\braket{\nu}{i}$.
Now, the Eq. (\ref{mirinv1}) can be written as
\begin{equation}
\label{mirinv2}
\sum_{\nu=0}^{N-1} e^{-i\varepsilon_{\nu} \tau}\ket{\nu} \braket{\nu}{i} = e^{i\phi_0}\sum_{\nu}(-1)^{\nu}\ket{\nu}\braket{\nu}{i}
\end{equation}

Therefore, perfect mirror inversion happens if for some time $\tau$,
\begin{equation}
\label{mirinv3}
\varepsilon_{\nu}\tau  = [2n(\nu) \pm \nu]\pi - \phi_0,
\end{equation}
where $n(\nu)$ is an arbitrary integer function.
%Analogous to the single-particle state, the $M$-particle state is defined as
%the one with $M$ sites occupied by fermions, or in the case of a spin chain, $M$-spins
%in state 1 and rest in state 0.
Since the spin system is mapped to a system of non-interacting spin-less fermions, if the mirror inversion happens for all the single-particle states, then mirror inversion also happens for all many-particle states up to a phase factor. This phase factor depends only on the number of particles (i.e., number of spins in state 1) in the many-particle sector and the total number of particles $N$.
Therefore, any spin chain whose Hamiltonian is symmetric and its single-particle eigenvalues satisfy Eq. (\ref{mirinv3}) generates mirror inversion of any input state up to a relative phase factor between its different subspaces which are spanned by different many-particle states. An example is given for a 5-spin chain in the later part of this section.
%Therefore, any spin chain whose Hamiltonian is symmetric and its single-particle eigenvalues satisfy Eq. (\ref{mirinv3}) generates mirror inversion of all single-particle as well as many-particle states up to a phase factor.

Apart from transferring the quantum state of a qubit to its mirror site, the mirror inversion operation in spin chains can also be used to transfer entangled states of multiple qubits to their mirror sites.

Alabanese \textit{et al}. \cite{Albanese} have considered a spin-chain with $J_i=\sqrt{i(N-i)}$, $h_i=0$, and whose
%We now consider one of the two chains, given in the reference \cite{Albanese} for which $J_i=\sqrt{i(N-i)}$ and $h_i=0$.
%For this chain, the 
$\cal{H}_1$ matrix is proportional to the $J_x$ rotation matrix of a spin $\frac{1}{2}(N-1)$ particle:
\begin{equation}
\cal{H}_1 = 2J_x = \sum_{i=1}^N \sqrt{i(N-i)}(\ket{n}\bra{n+1}+\ket{n+1}\bra{n}).
\end{equation}
The eigenvalues and eigenvectors of $J_x$ are well known \cite{Shankar}. Thus, the eigenvalues of $\cal{H}_1$ are $-(N-1), -(N-3), -(N-5) ~. . .~ (N-1)$. For time $\tau=\pi/2$, this spin chain does a perfect mirror inversion as it satisfies Eq. (\ref{mirinv3}). 

In the following, we first consider 4 and 5-spin XY chains of the above type and describe the decomposition of their
% discuss the cases of 4 and 5-spin XY chains of the above type and describe briefly the product decomposition of their
unitary evolution $U_{XY}(\frac{\pi}{2})=\exp (-i \cal{H}\frac{\pi}{2})$ into the Pauli operator basis using the algorithm of Ajoy \textit{et al}. \cite{Ashok}. 
%The proof and full details of the algorithm can be found in the original reference.  
Then, we generalize the decomposition to N-spin XY chains.

%We now describe briefly the product decomposition of the unitary evolution $U_{XY}(\frac{\pi}{2})=\exp (-i \cal{H}_{XY}\frac{\pi}{2})$ of the above chain for 4 and 5-spins

\subsection{4-spin chain}
The values of the nearest neighbour coupling constants for the 4-spin XY chain are $J_1=\sqrt{3}$, $J_2=2$, $J_3=\sqrt{3}$. For simplicity, we refer $U_{XY}(\frac{\pi}{2})$ as $U_{XY}$.
%and the evolution time for mirror inversion operation is $\tau=\frac{\pi}{2J}$.

%Now, we describe briefly the product decomposition of the 4-spin unitary operator $U_{XY}(\frac{\pi}{2J})=\exp (-i \cal{H}_{XY}\frac{\pi}{2J})$ into the Pauli operator basis using the algorithm of Ajoy \textit{et al}. []. The proof and full details of the algorithm can be found in the original reference. For simplicity, we refer $U_{XY}(\frac{\pi}{2J})$ as $U_{XY}$.

The Pauli operator basis for the $4$-spin system is given by,
\begin{equation}
B = \{\sigma_1^{\alpha} \sigma_2^{\beta} \sigma_3^{\gamma} \sigma_4^{\delta}\},
\end{equation}
where $\alpha, \beta, \gamma, \delta \in \{0, x, y, z\}$ and $\sigma^0 = \mathbbm{1}$.
%\begin{equation}
%B=\{\mathbbm{1}, ~\sigma_i^{\alpha}, ~\sigma_i^\alpha \sigma_j^\beta, ~\sigma_i^\alpha \sigma_j^\beta \sigma_l^\gamma, ~\sigma_i^\alpha \sigma_j^\beta \sigma_l^\gamma \sigma_n^\delta \},
%\end{equation}
%where $\{i,j,l,n\} \in \{1,2,3,4\}$, $\{\alpha, \beta, \gamma, \delta\} \in \{x,y,z\}$ for example, $\sigma_2^x=\mathbbm{1} \otimes \sigma^x \otimes \mathbbm{1} \otimes \mathbbm{1}$.
Our aim is to product decompose $U_{XY}$ into the Pauli operator basis as follows,
\begin{equation}
U_{XY}=\prod_{k=1}^m \exp (-i \theta_k D_k); ~~~D_k \in B.
\label{prod}
\end{equation}
Since $B$ forms a complete basis, $U_{XY}$ can be expanded as a sum in $B$ as follows,
\begin{equation}
U_{XY} = \tfrac{1}{4} \sum_{\alpha,\beta} \eta_{\alpha\beta} ~(\sigma_1^{\alpha} \sigma_2^{\beta} \sigma_3^{\beta} \sigma_4^{\alpha})
\label{uxysum}
\end{equation}
where $\alpha, \beta \in \{0, x, y, z\}$. The coefficient $\eta_{\alpha \beta}=i$ if 
(i) $\alpha \not= \beta$ and (ii) either of $\alpha \in \{0,z\}$ or $\beta \in \{0,z\}$, but not both. For all other cases $\eta_{\alpha \beta}=1$.

%\begin{align}
%U_{XY} = & \, \tfrac{1}{4} \big[(\mathbbm{1} + \sigma_2^z \sigma_3^z + \sigma_1^z \sigma_4^z + \sigma_1^x \sigma_2^x \sigma_3^x \sigma_4^x + \sigma_1^y  \sigma_2^y \sigma_3^y \sigma_4^y \nonumber \\
%& + \sigma_1^z  \sigma_2^z \sigma_3^z \sigma_4^z + \sigma_1^x  \sigma_2^y \sigma_3^y \sigma_4^x   
%+ \sigma_1^y \sigma_2^x \sigma_3^x \sigma_4^y) \nonumber \\
%& + i (\sigma_2^x \sigma_3^x + \sigma_2^y \sigma_3^y
%+ \sigma_1^y \sigma_4^y + \sigma_1^x \sigma_4^x + \sigma_1^x \sigma_2^z \sigma_3^z \sigma_4^x \nonumber \\
%& + \sigma_1^y \sigma_2^z \sigma_3^z \sigma_4^y
%+ \sigma_1^z \sigma_2^x \sigma_3^x \sigma_4^z + \sigma_1^z \sigma_2^y \sigma_3^y \sigma_4^z)\big].
%\end{align}
\begin{figure} [t]
 \centering
 \includegraphics[width=6cm]{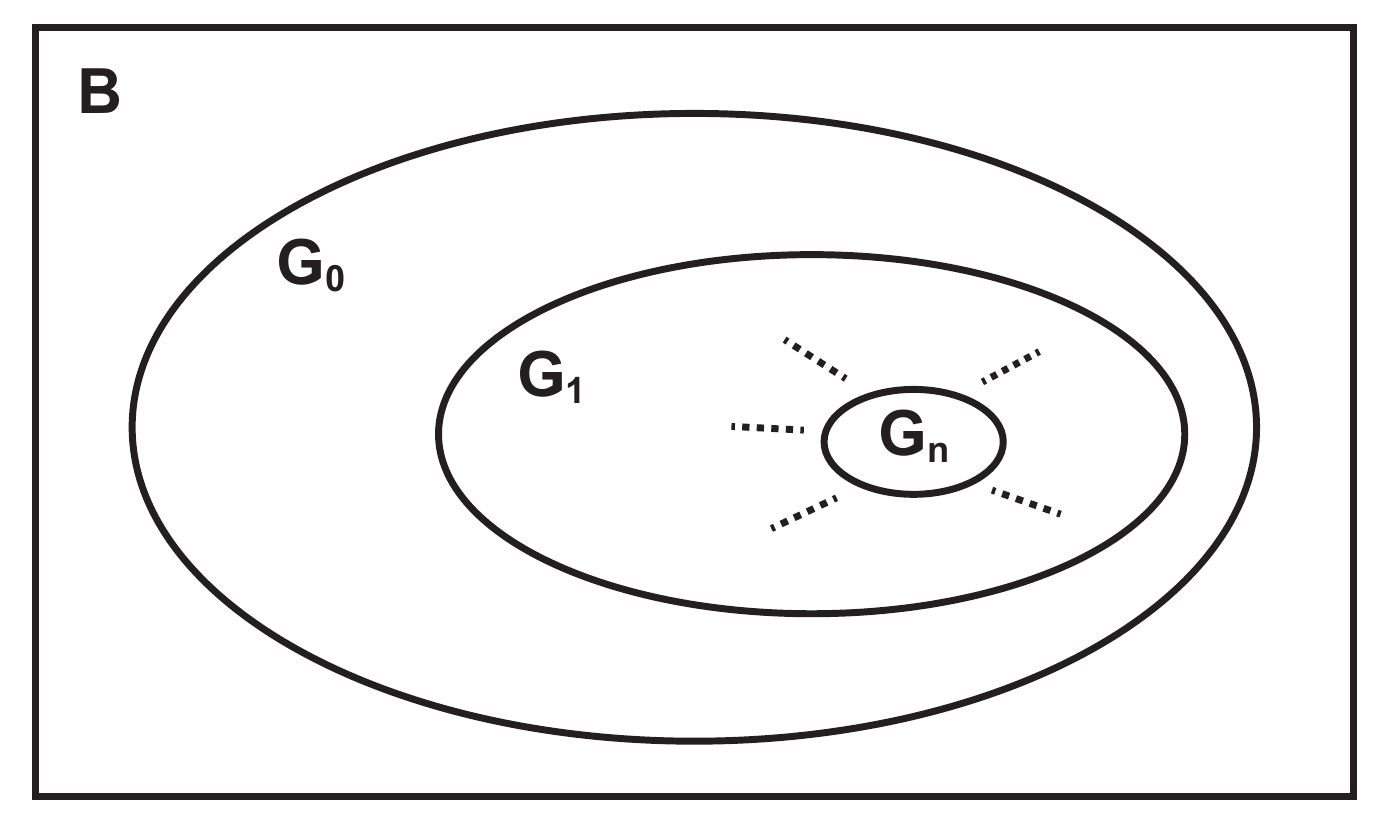}
 \caption{The progressive reduction of search-space by
 the decomposition algorithm.  Here $B$ is the full Pauli operator basis,
 and $G_0$, $G_1$,...,$G_n$ are recursive subgroups ($G_{k+1} \subset G_k$).
The algorithm proceeds till one obtains $G_n=\{\mathbbm{1}\}$.}
 \label{fig0} 
\end{figure}

The product decomposition algorithm \cite{Ashok} proceeds through a systematic reduction of the search space, which is shown schematically in Fig. (\ref{fig0}).
Let us consider the set,
\begin{align}
G_0 = \{& \mathbbm{1}, \sigma_2^x \sigma_3^x, \sigma_2^y \sigma_3^y, \sigma_2^z \sigma_3^z, \sigma_1^x \sigma_4^x,
\sigma_1^y \sigma_4^y, \sigma_1^z \sigma_4^z, \nonumber \\
& \sigma_1^x \sigma_2^x \sigma_3^x \sigma_4^x, \sigma_1^y \sigma_2^y \sigma_3^y \sigma_4^y, 
\sigma_1^z \sigma_2^z \sigma_3^z \sigma_4^z, \nonumber \\
& \sigma_1^x \sigma_2^y \sigma_3^y \sigma_4^x, \sigma_1^y \sigma_2^x \sigma_3^x \sigma_4^y, 
\sigma_1^x \sigma_2^z \sigma_3^z \sigma_4^x, \nonumber \\
& \sigma_1^z \sigma_2^x \sigma_3^x  \sigma_4^z,
\sigma_1^y \sigma_2^z \sigma_3^z \sigma_4^y, \sigma_1^z \sigma_2^y \sigma_3^y \sigma_4^z \},
\end{align}
which contains all the elements from the sum expansion in Eq. (\ref{uxysum}). This set $G_0$ forms a group under multiplication of operators. This implies, all the operators $D_k$ in the product expansion of Eq. (\ref{prod}) belong exclusively to $G_0$. In cases, where $G_0$ doesn't form a group by itself, one can add minimum number of operators from $B$ to $G_0$ so that $G_0$ forms a group.
Now, consider the set,
\begin{align}
G_1 = \{& \mathbbm{1}, \sigma_2^x \sigma_3^x, \sigma_2^y \sigma_3^y, \sigma_2^z \sigma_3^z, \sigma_1^x \sigma_2^x \sigma_3^x \sigma_4^x, \sigma_1^x \sigma_2^y \sigma_3^y \sigma_4^x, \nonumber \\
& \sigma_1^x \sigma_2^z \sigma_3^z \sigma_4^x, \sigma_1^x \sigma_4^x \},
\end{align}
which is a subgroup of $G_0$. The set $G_1$ is selected such that it is the biggest possible subgroup of $G_0$. However, this selection need not be unique.

There are a total number of $m$ operators ($D_k$ s) in the product decomposition of Eq. (\ref{prod}) and as said earlier, all of them belong to $G_0$. Let us suppose that $m'$ of these operators belong to ($G_0-G_1$). Then,
the next and key step of the algorithm is to find out these $m'$ operators $D_k \in (G_0-G_1)$ and the corresponding angles $\theta_k$, such that $U_{XY} \prod_{k=1}^{m'} \exp (i \theta_k D_k)$ can be expanded as a sum in $B$, whose elements belong exclusively to $G_1$. i.e.,
\begin{align}
U_{XY} \prod_{k=1}^{m'} \exp (i \theta_k D_k) &= U_{XY}^{(m')} \nonumber \\
&= \sum_r \frac{1}{\tau} ~\mathrm{Tr}(U_{XY}^{(m')} \tilde{D}_r^{\dagger}) ~\tilde{D}_r,
\end{align}
where $\tilde{D}_r \in G_1$, and $\tau=\mathrm{Tr}(\tilde{D}_r^\dagger \tilde{D}_r)$.

We define the norm of the space spanned by the elements of a set $G$ in a unitary $U$ as,
\begin{equation}
N_G(U)=\sum_n \vert \mathrm{Tr}(UD_n^\dagger)/\mathrm{Tr}(D_n^\dagger D_n)\vert^2; ~~~D_n \in G.
\end{equation}
The intuitive meaning of norm $N_G(U)$ is - to what extent $U$ can be constructed using the elements of $G$.
%In order to maximize this norm, 

For a chosen operator $D_k$, the angle $\theta_k$ is chosen such that 
the quantity
$N_{G_1} (U_{XY}^{(k)} ) - N_{G_1}(U_{XY}^{(k-1)})$ is maximized. This leads to, 
\begin{equation}
\theta_{k} = \frac{1}{2} \tan^{-1}\left(\frac{W_{k-1}(D_k)}{\Delta_{k-1}}\right), 
\end{equation}
where,
\begin{align}
W_q(D) &= \frac{1}{\tau^2} \mathrm{Im} \sum_r 
%\left\{ 
\mathrm{Tr} \left( U_{XY}^{(q)} \tilde{D}_r^\dagger\right) 
\mathrm{Tr} \left(U_{XY}^{(q)}  D^\dagger \tilde{D}_r^\dagger \right)^*,
%\right\};
% ~\tilde{D}_r \in G_1, 
\nonumber \\
\Delta_q &= N_{G_1} \left( U_{XY}^{(q)}\right)  - \frac{1}{2}. \nonumber
\end{align}

The next operator $D_{k+1}$ can be chosen as follows. Calculate the quantity 
%\begin{align}
%%W'(D)&= -\frac{1}{\tau^2} \mathrm{Re} \sum_r i \Bigg\{ \mathrm{Tr}\bigg(U_{XY}\prod_{p=1}^{k} \exp (i \theta_p D_p) ~\tilde{D}_r^\dagger\bigg) \nonumber  \\
%%&\times \mathrm{Tr}\bigg(U_{XY}\prod_{p=1}^{k} \exp (i \theta_p D_p) ~D^\dagger \tilde{D}_r^\dagger\bigg)^* \Bigg\}; ~\tilde{D}_r \in G_1, \nonumber \\
%W'(D) &= \frac{1}{\tau^2} \mathrm{Im} \sum_r 
%%\left\{ 
%\mathrm{Tr} \left( U_{XY}^{(k)} \tilde{D}_r^\dagger\right) 
%\mathrm{Tr} \left(U_{XY}^{(k)}  D^\dagger \tilde{D}_r^\dagger \right)^*,
%\end{align}
$W_k(D)$ for all the operators $D$ in $(G_0-G1)$. Choose the operator $D$ that maximizes $W_k(D)$ as the operator $D_{k+1}$.

By following the above procedure, we get,
\begin{align}
U_{XY} & \exp(-i \tfrac{\pi}{4} \sigma_1^y \sigma_2^z \sigma_3^z \sigma_4^y) = \nonumber \\
& \quad \tfrac{1}{\sqrt{8}} \big[(\mathbbm{1} + \sigma_2^z \sigma_3^z + \sigma_1^x \sigma_2^x \sigma_3^x \sigma_4^x + \sigma_1^x \sigma_2^y \sigma_3^y \sigma_4^x) \nonumber \\ 
& \quad + i (\sigma_2^x \sigma_3^x + \sigma_2^y \sigma_3^y  + \sigma_1^x \sigma_4^x + \sigma_1^x \sigma_2^z \sigma_3^z \sigma_4^x )\big],
\end{align}
where all the operators on the right hand side belong to $G_1$. Now consider the set,
\begin{equation}
G_2 = \{\mathbbm{1}, \sigma_2^z \sigma_3^z,
\sigma_2^x \sigma_3^x, \sigma_2^y \sigma_3^y \},
\end{equation}
which is a subgroup of $G_1$. By repeating the above procedure, we get,
\begin{align}
U_{XY} \exp(-i \tfrac{\pi}{4} \sigma_1^y \sigma_2^z \sigma_3^z \sigma_4^y)  \exp(-i \tfrac{\pi}{4} \sigma_1^x \sigma_2^z \sigma_3^z \sigma_4^x) = \nonumber \\
\tfrac{1}{2} \big[\mathbbm{1} + \sigma_2^z \sigma_3^z  
+ i (\sigma_2^x \sigma_3^x + \sigma_2^y \sigma_3^y)\big],
\end{align}
where again all the operators on the right hand side belong to $G_2$.

This process is repeated till we get $G_n=\{\mathbbm{1}\}$. Here, by repeating the above procedure two more times with $G_3 = \{\mathbbm{1}, \sigma_2^x \sigma_3^x \}$ and $G_4=\{\mathbbm{1}\}$, we get the full product decomposition as
\begin{align}
U_{XY} \exp(-i \tfrac{\pi}{4} \sigma_1^y \sigma_2^z \sigma_3^z \sigma_4^y) \exp(-i \tfrac{\pi}{4} \sigma_1^x \sigma_2^z \sigma_3^z \sigma_4^x) & \nonumber \\
\times \exp(-i \tfrac{\pi}{4} \sigma_2^x \sigma_3^x) \exp(-i \tfrac{\pi}{4} \sigma_2^y \sigma_3^y) &
= \mathbbm{1},
\end{align}
which can be written as,
\begin{align}
\label{Uop4}
U_{XY} = \, & \exp(i \tfrac{\pi}{4} \sigma_1^y \sigma_2^z \sigma_3^z \sigma_4^y) \exp(i \tfrac{\pi}{4} \sigma_1^x \sigma_2^z \sigma_3^z \sigma_4^x) \nonumber \\
& \times \exp(i \tfrac{\pi}{4} \sigma_2^x \sigma_3^x) \exp(i \tfrac{\pi}{4} \sigma_2^y \sigma_3^y).
\end{align}

%By using the algorithm given by Ajoy \textit{et al}. \cite{Ashok}, the 4-spin unitary operator $U_{XY}(\tfrac{\pi}{2J})$ can be decomposed into the Pauli operator basis as follows,
%\begin{eqnarray}
%\label{Uop4}
%U_{XY}(\tfrac{\pi}{2J})= \ &&\exp(i\tfrac{\pi}{4}\sigma_1^x \sigma_2^z \sigma_3^z \sigma_4^x) \exp(i\tfrac{\pi}{4}\sigma_1^y \sigma_2^z \sigma_3^z \sigma_4^y ) \nonumber\\ && \times
%\exp(i\tfrac{\pi}{4}\sigma_2^x \sigma_3^x) \exp(i\tfrac{\pi}{4}\sigma_2^y \sigma_3^y).
%\end{eqnarray}\\

\subsection{5-spin chain}
The values of the nearest neighbour coupling constants for the 5-spin XY chain are $J_1=2$, $J_2=\sqrt{6}$, $J_3=\sqrt{6}$, $J_4=2$.
%and the evolution time for mirror inversion operation is $\tau=\frac{\pi}{2J}$. 
The unitary evolution of this 5-spin chain for a time $\tau=\pi/2$, i.e., $U_{XY}(\frac{\pi}{2})=\exp (-i \cal{H} \frac{\pi}{2})$, produces the mirror image of any 5-spin input state up to a phase difference. For example,
%This means that the application of the unitary operator $U_{XY}(\frac{\pi}{2})=\exp (-i \cal{H}_{XY}\frac{\pi}{2})$ on any 5-qubit input state produces its mirror image up to a phase difference. For example,
\begin{eqnarray}
U_{XY}(\tfrac{\pi}{2}) \tfrac{1}{\sqrt{2}} (\ket{00}+\ket{11})_{12} \ket{000}_{345} = \nonumber \\
\ket{000}_{123} \tfrac{1}{\sqrt{2}} (\ket{00}-\ket{11})_{45} ~~\mathrm{and}
\end{eqnarray}
\begin{eqnarray}
U_{XY}(\tfrac{\pi}{2}) \tfrac{1}{\sqrt{2}} (\ket{01}+\ket{10})_{12} \ket{000}_{345} = \nonumber \\ 
\ket{000}_{123} \tfrac{1}{\sqrt{2}} (\ket{01}+\ket{10})_{45}.
\end{eqnarray}
The above equations show that entangled states can be transferred from one end of the chain to the other up to a phase difference.

By following the procedure similar to that of the 4-spin case, the 5-spin unitary operator $U_{XY}(\tfrac{\pi}{2})$ is decomposed into the Pauli operator basis and it is given by,
%The decomposition of the 5-spin unitary operator $U_{XY}(\tfrac{\pi}{2J})$ in to the Pauli operator basis is given by
\begin{align}
\label{Uop}
U_{XY}(\tfrac{\pi}{2})=&\exp(i\tfrac{\pi}{4}\sigma_1^x \sigma_2^z \sigma_3^z \sigma_4^z \sigma_5^y) 
\exp(i\tfrac{\pi}{4}\sigma_1^y \sigma_2^z \sigma_3^z \sigma_4^z \sigma_5^x) \nonumber\\
&\times \exp(i\tfrac{\pi}{4}\sigma_2^x \sigma_3^z \sigma_4^y)
\exp(i\tfrac{\pi}{4} \sigma_2^y \sigma_3^z \sigma_4^x) \nonumber\\
&\times \exp(i\tfrac{\pi}{2}\sigma_1^x \sigma_2^y \sigma_4^y \sigma_5^x)
\end{align}

\subsection{$N$-spin chain}
The above decomposition can be extended to $N$-spin XY chains, which is given as follows: 

When $N$ is odd,
\begin{align}
\label{Uoddn}
U_{XY}(\tfrac{\pi}{2})&= \exp(\pm i\tfrac{\pi}{4}\sigma_1^x \sigma_2^z \sigma_3^z \cdots \sigma_{N-1}^z \sigma_N^y) \nonumber \\
& \exp(\pm i\tfrac{\pi}{4}\sigma_1^y \sigma_2^z \sigma_3^z \cdots \sigma_{N-1}^z \sigma_N^x) \nonumber \\
& \exp(\pm i\tfrac{\pi}{4}\sigma_2^x \sigma_3^z \sigma_4^z \cdots \sigma_{N-2}^z \sigma_{N-1}^y) \nonumber \\
& \exp(\pm i\tfrac{\pi}{4} \sigma_2^y \sigma_3^z \sigma_4^z \cdots \sigma_{N-2}^z \sigma_{N-1}^x) ~\cdots \nonumber \\
& \exp(\pm i\tfrac{\pi}{4} \sigma_{\frac{N-1}{2}}^x \sigma_{\frac{N+1}{2}}^z \sigma_{\frac{N+3}{2}}^y)\nonumber \\
& \exp(\pm i\tfrac{\pi}{4} \sigma_{\frac{N-1}{2}}^y \sigma_{\frac{N+1}{2}}^z \sigma_{\frac{N+3}{2}}^x) \nonumber \\
& \exp(\pm i\tfrac{\pi}{2}\sigma_1^x \sigma_2^y \sigma_3^x \cdots \mathbbm{1}_{\frac{N+1}{2}} \cdots \sigma_{N-2}^x \sigma_{N-1}^y \sigma_{N}^x),
\end{align}
where the signs $+$ and $-$ are for chains having $4m+1$ and $4m+3$ ($m$ is an integer) number of spins respectively and $\mathbbm{1}_{\frac{N+1}{2}}$ is the identity operator for spin $\frac{N+1}{2}$.

When $N$ is even,
\begin{align}
\label{Uevenn}
U_{XY}(\tfrac{\pi}{2})&= \exp(\pm i\tfrac{\pi}{4}\sigma_1^x \sigma_2^z \sigma_3^z \cdots \sigma_{N-1}^z \sigma_N^x) \nonumber \\
& \exp(\pm i\tfrac{\pi}{4}\sigma_1^y \sigma_2^z \sigma_3^z \cdots \sigma_{N-1}^z \sigma_N^y) \nonumber \\
& \exp(\pm i\tfrac{\pi}{4}\sigma_2^x \sigma_3^z \sigma_4^z \cdots \sigma_{N-2}^z \sigma_{N-1}^x) \nonumber \\
& \exp(\pm i\tfrac{\pi}{4} \sigma_2^y \sigma_3^z \sigma_4^z \cdots \sigma_{N-2}^z \sigma_{N-1}^y) ~\cdots \nonumber \\
& \exp(\pm i\tfrac{\pi}{4} \sigma_{\frac{N}{2}}^x \sigma_{\frac{N}{2}+1}^x)\nonumber \\
& \exp(\pm i\tfrac{\pi}{4} \sigma_{\frac{N}{2}}^y \sigma_{\frac{N}{2}+1}^y),
\end{align}
where the signs $+$ and $-$ are for chains having $4m$ and $4m+2$ ($m$ is an integer) number of spins respectively.

The unitary operators in the right hand side of Eqs. (\ref{Uop4}) and (\ref{Uop}) can be further decomposed into single qubit rotations and two qubit gates \cite{Tseng}. 
%The full decomposition of each of these operators is given in the Appendix. 
Conventional pulse sequences can be constructed by using these decompositions to realize the full unitary evolution. However, in the experimental implementation, we realized each of these operators with a single GRAPE pulse.
We now describe the experimental simulation of the above $4$ and $5$-spin chains using NMR techniques.

\section{Experimental Implementation}
\label{sec-results}

\begin{figure}
 \centering
 \includegraphics[width=8cm]{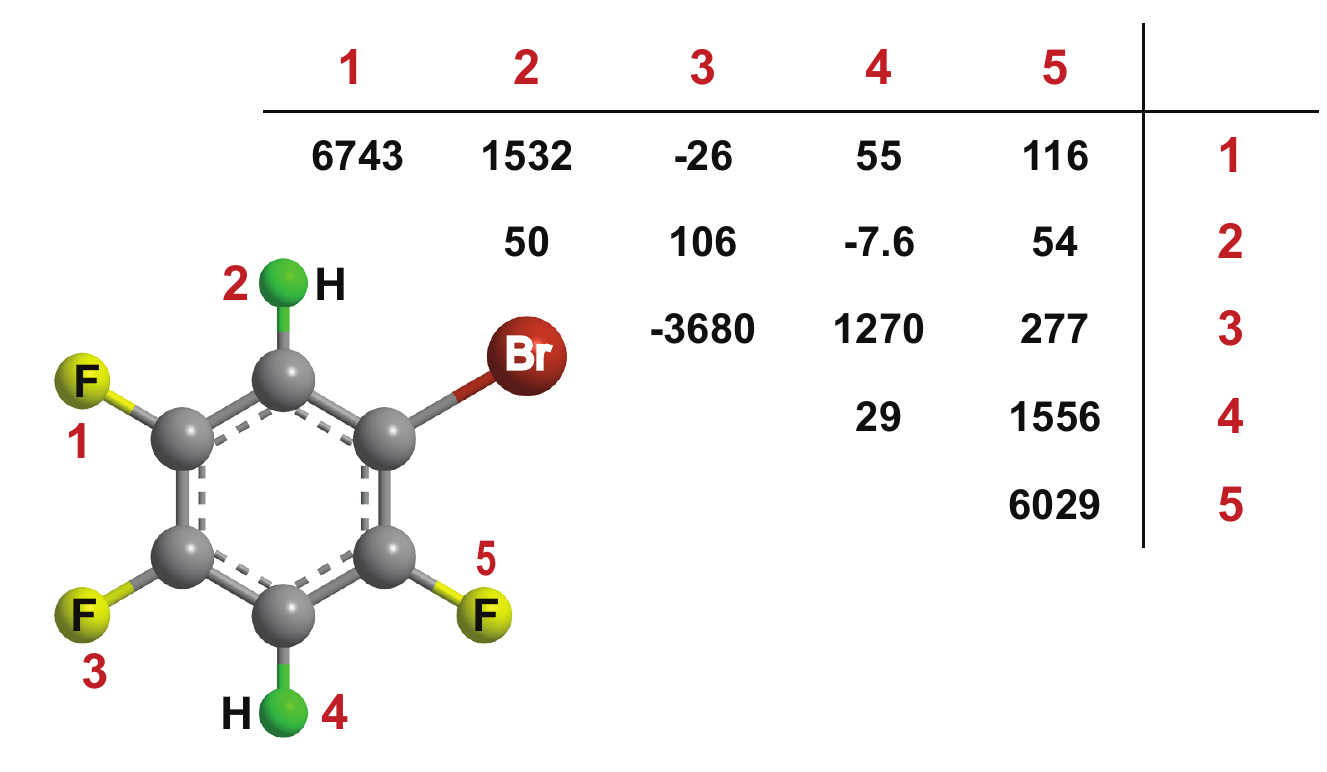}
 \caption{Chemical structure of the molecule and Hamiltonian parameters. In the table, diagonal elements correspond to the Zeeman shifts ($\nu_i$) of the nuclear spins (in Hz) in the doubly rotating frame and the off-diagonal elements correspond to the coupling constants ($J_{ij}+2D_{ij}$) between them (in Hz).}
 \label{fig1} 
\end{figure}

We choose 1-bromo-2,4,5-trifluorobenzene partially oriented in a liquid crystal medium, 
N-(4-methoxybenzaldehyde)-4-butylaniline (MBBA) as our spin system for the experimental implementation. 
The three $^{19}$F and two $^1$H nuclei form a 5-spin system. These spins are labelled as $1$, $2$, $3$, $4$, and $5$ as shown in the Fig. \ref{fig1}.  The effective transverse relaxation times ($T_2^*$) of the transitions of spins $1$, $3$, and $5$ (fluorine nuclei) are in the range 40-60 ms, 40-60 ms, and 60-100 ms and that of the transitions of spins $2$ and $4$ (proton nuclei) are in the range 140-150 ms, and 110-150 ms, respectively. 
These are calculated from the inverse of the observed line-widths, which are mainly governed by the director fluctuation in a liquid crystal and are an order of magnitude larger than
the line-widths in isotropic solutions.
All the experiments have been carried out at an ambient temperature of 300 K on a Bruker AVIII 500 MHz NMR spectrometer using a QXI probe.

Due to the partial orientational order of the liquid crystal medium, the direct dipolar couplings among the spins do not get fully averaged out, but get scaled down by the order parameter. The residual dipolar coupling between the spins is an order of magnitude stronger than the indirect scalar coupling. The Zeeman shift values of the nuclear spins and the coupling constants between them are given in the Fig. \ref{fig1}. The Hamiltonian for the dipolar interaction between the hetero-nuclear spins is of the form $\cal{H}_D=\frac{\pi}{2} \sum_{i,j(i<j)} D_{ij} 2\sigma_i^z \sigma_j^z$, where $D_{ij}$ is the scaled dipolar coupling constant, and the same between the homo-nuclear spins is of the form $\cal{H}_D=\frac{\pi}{2} \sum_{i,j(i<j)} D_{ij} (3 \sigma_i^z \sigma_j^z - \sigma_i\cdot\sigma_j)$. Since the difference between Zeeman shifts of any pair of spins is much larger than the respective dipolar coupling between them, the Hamiltonian for the homo-nuclear dipolar interaction can be approximated to $\cal{H}_D=\frac{\pi}{2} \sum_{i,j(i<j)} D_{ij} 2\sigma_i^z \sigma_j^z$. Hence, the full Hamiltonian of the spin system in the doubly rotating frame can be written as
\begin{equation}
\cal{H}_{NMR} = -\pi \sum_i \nu_i \sigma_i^z + \frac{\pi}{2} \sum_{i,j(i<j)} (J_{ij}+2D_{ij}) \sigma_i^z \sigma_j^z,
\end{equation}
where $\nu_i$ is the Zeeman shift of the spin $i$ and $J_{ij}$ is the scalar coupling constant between the spins $i$ and $j$. The magnitude of the coupling constants $(J_{ij}+2D_{ij})$ was obtained by fitting equilibrium spectra of the spin system and sign of them was determined by using hetero-nuclear Z-COSY experiments \cite{ZCOSY, Grace, RanaZCOSY}. 

The equilibrium deviation density matrix of the spin system under high temperature and high field approximation can be represented by \cite{Ernst},
\begin{equation}
\rho_{eq}^\Delta = \gamma_\mathrm{F} (\sigma_z^1 + \sigma_z^3 + \sigma_z^5) + \gamma_\mathrm{H} (\sigma_z^2 + \sigma_z^4),
\end{equation}
where $\gamma_\mathrm{H}$ and $\gamma_\mathrm{F}=0.94\gamma_\mathrm{H}$ are gyromagnetic ratios of the nuclei $^{19}$F and $^{1}$H respectively.

The 5-spin NMR system is used to demonstrate the mirror inversion operation in the following XY chains: (i) a 5-spin chain, prepared in mixed or subsystem pseudo-pure initial states and (ii) a 4-spin chain prepared in pseudo-pure initial states.
We used GRAPE \cite{GRAPE} technique to realize the product decompositions of the unitary evolutions of $4$ and $5$ spin XY chains which are given in Eqs. (\ref{Uop4}) and (\ref{Uop}).
Each of the unitary operators in the right hand side of these equations has been realized using a single GRAPE pulse. The total length of these pulses for simulating 4-spin chain is 34 ms and that for simulating 5-spin chain is 40 ms.
All the GRAPE pulses were optimized such that they are robust against RF field inhomogeneity and the average 
Hilbert-Schmidt fidelity of all these pulses are greater than 0.99.

\subsection{Five-spin initial states}
\begin{figure}
\hspace{-0.43cm}
 \centering
 \includegraphics[width=9cm]{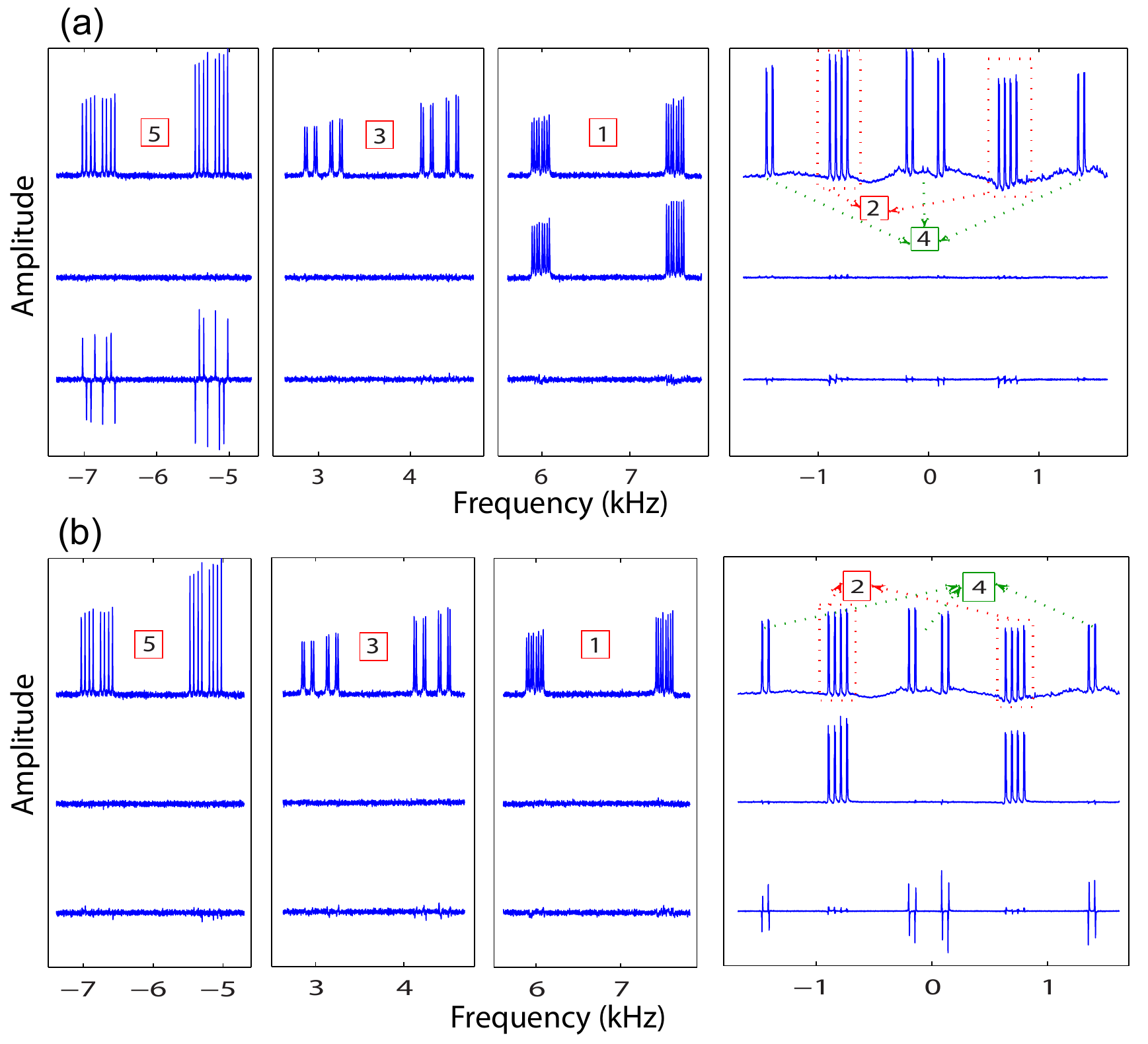}
 \caption{Experimental results for quantum state transfer. (a) Quantum state transfer of spin $1$ to its mirror image $5$ and (b) quantum state transfer of spin $2$ to its mirror image $4$. In both (a) and (b) the top row correspond to the equilibrium spectra, the middle row represent spectra corresponding to the initial state, and the bottom row is that of the final state.}
 \label{fig2} 
\end{figure}

\emph{Quantum state transfer:} As described in the previous section, mirror inversion operation can be used to transfer quantum state of a spin to its mirror image. Here, we perform two different experiments with respective initial states (i) $\sigma_1^x$ and (ii) $\sigma_2^x$. These initial states were prepared from the equilibrium state as follows. We first apply a spin selective $(\pi/2)^x$ pulse on spin $1$ ($2$) and then a $(\pi/2)^{-x}$  pulse on all the spins followed by a gradient pulse in the z-direction. This saturates the magnetization of all the spins except spin $1$ ($2$). We now apply a spin selective $(\pi/2)^y$ pulse on spin $1$ ($2$) which produces the desired initial state $\sigma_1^x$ ($\sigma_2^x$). 
All the spin selective and global pulses used here and hence forth were realized using GRAPE technique \cite{GRAPE}.
The lengths of the spin selective pulses on fluorine spins ($1$, $3$, and $5$) are in the range 500-600 $\mu$s and those on the proton spins $2$ and $4$ are 2.5 ms and 3 ms respectively. The length of the $\pi/2$ pulse on all the spins is 500 $\mu$s.
%All these pulses were realized by using GRAPE technique \cite{GRAPE}. 
The resultant spectra which confirm the creation of the initial states are shown in the middle trace of Figs. \ref{fig2}(a) and \ref{fig2}(b). The application of the unitary operator in Eq. (\ref{Uop}) on the initial states $\sigma_1^x$ and $\sigma_2^x$ produces the states $\sigma_1^z \sigma_2^z \sigma_3^z \sigma_4^z \sigma_5^x$ and $\sigma_1^z \sigma_2^z \sigma_3^z \sigma_4^x \sigma_5^z$ respectively. This clearly demonstrates that the coherence of spin $1$ ($2$) is transferred to its mirror image $5$ ($4$). The transferred coherence is anti-phase with respect to all other spins, this is due to the relative phase difference between different many-particle subspaces, which is explained in the previous section. The experimental spectra corresponding to the final states are shown in the bottom trace of Figs. \ref{fig2}(a) and \ref{fig2}(b). The clear anti-phase signals for the spins $5$ and $4$ and the absence of signals for all the other spins indicate the efficient implementation of mirror inversion operation.

\emph{Entanglement transfer:} As discussed in the previous section, mirror inversion operation can also be used to transfer entangled states from one end of the chain to the other. We initially prepare maximally entangled states of spins $1$ and $2$. The unitary evolution of the spin chain drives these entangled states to their mirror images i.e., spins $4$ and $5$. We choose the initial states as
$\tfrac{1}{{16}} (\ket{00}+\ket{11})(\bra{00}+\bra{11})_{12} \otimes \mathbbm{1}_3 \otimes \mathbbm{1}_4 \otimes \mathbbm{1}_5$ and $\tfrac{1}{{16}} (\ket{01}+\ket{10})(\bra{01}+\bra{10})_{12} \otimes \mathbbm{1}_3 \otimes \mathbbm{1}_4 \otimes \mathbbm{1}_5$, where spins $1$ and $2$ are in the maximally entangled state and all other spins are in the maximally mixed state. These initial states were prepared from the equilibrium as described below. We first prepare the spins $1$ and $2$ in the $\ket{00}$ pseudo-pure state using spatial averaging technique \cite{PPS1} and all the other spins in the maximally mixed state by dephasing their magnetization. The pulse sequence, used to create the state $\tfrac{1}{{8}} \ket{00}\bra{00}_{12} \otimes \mathbbm{1}_3 \otimes \mathbbm{1}_4 \otimes \mathbbm{1}_5$ from equilibrium is given by
\begin{eqnarray}
\bigg[\frac{\pi}{2}\bigg]_x^{1,2} - \bigg[\frac{\pi}{2}\bigg]_{-x}^{All} - \bigg[G\bigg]_z - \bigg[0.32\pi \bigg]_x^1 - 
\bigg[G\bigg]_z \nonumber \\
- \bigg[\frac{\pi}{4}\bigg]_x^{2} - \bigg[\frac{1}{2(J+2D)_{12}}\bigg] - \bigg[\frac{\pi}{4}\bigg]_{-y}^{2} -
\bigg[G\bigg]_z,
\end{eqnarray}
where $[\theta]_{\alpha}^i$ denotes a $\theta$ degree pulse on spin $i$ about the axis $\alpha$, $[G]_z$, a gradient pulse along z-direction, and $[\frac{1}{2(J+2D)_{ij}}]$ represents coupling evolution of spins $i$,$j$ for a period $\frac{1}{2(J+2D)_{ij}}$.
Here, the two $\pi/4$ pulses on spin $2$ and the coupling evolution in between were combined and realized using a single GRAPE pulse. The length of this pulse is 2.4 ms.
The state $\tfrac{1}{{8}} \ket{10}\bra{10}_{12} \otimes \mathbbm{1}_3 \otimes \mathbbm{1}_4 \otimes \mathbbm{1}_5$ can be prepared from the state $\tfrac{1}{{8}} \ket{00}\bra{00}_{12} \otimes \mathbbm{1}_3 \otimes \mathbbm{1}_4 \otimes \mathbbm{1}_5$ by applying a $\pi$ pulse on spin $1$.
%The above unitary operators were combined where possible (operators between the gradients) and the resulting operators were realized using GRAPE pulses.
The desired initial states $\tfrac{1}{{16}} (\ket{00}+\ket{11})(\bra{00}+\bra{11})_{12} \otimes \mathbbm{1}_3 \otimes \mathbbm{1}_4 \otimes \mathbbm{1}_5$ and $\tfrac{1}{{16}} (\ket{01}+\ket{10})(\bra{01}+\bra{10})_{12} \otimes \mathbbm{1}_3 \otimes \mathbbm{1}_4 \otimes \mathbbm{1}_5$ were prepared by applying the unitary operator $e^{-i\frac{\pi}{4}\sigma_1^x \sigma_2^y}$ on the states $\ket{00}\bra{00}_{12} \otimes \mathbbm{1}_3 \otimes \mathbbm{1}_4 \otimes \mathbbm{1}_5$ and $\ket{10}\bra{10}_{12} \otimes \mathbbm{1}_3 \otimes \mathbbm{1}_4 \otimes \mathbbm{1}_5$ respectively. This unitary operator was also realized using a GRAPE pulse and its length is 2.4 ms. \\

Applying the mirror inversion operator in Eq. (\ref{Uop}) on the above initial states leads to:
\begin{eqnarray}
& \hspace{-0.8cm}  \tfrac{1}{{16}} (\ket{00}+\ket{11})(\bra{00}+\bra{11})_{12} \otimes \mathbbm{1}_3 \otimes \mathbbm{1}_4 \otimes \mathbbm{1}_5 \longrightarrow \nonumber \\ 
& \tfrac{1}{{16}} \mathbbm{1}_1 \otimes \mathbbm{1}_2 \otimes \mathbbm{1}_3 \otimes (\ket{00}-\ket{11})(\bra{00}-\bra{11})_{45},
\end{eqnarray}
\begin{eqnarray}
& \hspace{-0.8cm}  \tfrac{1}{{16}} (\ket{01}+\ket{10})(\bra{01}+\bra{10})_{12} \otimes \mathbbm{1}_3 \otimes \mathbbm{1}_4 \otimes \mathbbm{1}_5 \longrightarrow \nonumber \\ 
& \tfrac{1}{{16}} \mathbbm{1}_1 \otimes \mathbbm{1}_2 \otimes \mathbbm{1}_3 \otimes (\ket{01}+\ket{10})(\bra{01}+\bra{10})_{45},
\end{eqnarray}
where now spins $1$, $2$, and $3$ are in the maximally mixed state and spins $4$, and $5$ are in the maximally entangled state.

To confirm the creation and transfer of entangled states, we have performed quantum state tomography of $2$-spin subsystems containing spins $1$, $2$ and $4$, $5$ for both initial and final states. The procedure is described below \cite{Avik2005}.

The $2$-spin density matrix contains $6$ complex off-diagonal ($4$ single quantum, $1$ double quantum, and $1$ zero quantum) and $3$ real diagonal, independent elements. Out of these, only single quantum elements are directly observable. To reconstruct the full density matrix, one needs to measure the other elements by converting them into single quantum. The diagonal elements are measured by applying a gradient pulse, followed by a $\pi/2$ on each spin separately. Although, all the single quantum elements can be measured at a time directly, for proper scaling, we measured each one of them separately. The single quantum elements of the $i^\mathrm{th}$ spin are observed by using the following pulse sequences,
\begin{align}
A. ~&\bigg[\frac{\pi}{2}\bigg]_\alpha^i - \bigg[G\bigg]_z - \bigg[\frac{\pi}{2}\bigg]_y^i, \\
B. ~&\bigg[\frac{\pi}{2}\bigg]_\alpha^i - \bigg[G\bigg]_z -\bigg[\pi\bigg]^j \bigg[\frac{\pi}{2}\bigg]_y^i,
\end{align}
where $i$, $j$ represent the spins which are being measured. The experiments $A$ and $B$ are repeated twice, with the phase $\alpha$ as $-y$ and $x$ for measuring the real and imaginary parts respectively. Here, the first $\pi/2$ pulse followed by a gradient dephases all the elements except the selected single quantum elements which are converted into diagonal part. The final $\pi/2$ pulse converts these diagonal elements back into real single quantum elements, which are measured. The same pulse sequence can be used for measuring the single quantum elements of $j^\mathrm{th}$ spin, by replacing pulses on $i^\mathrm{th}$ spin with pulses on $j^\mathrm{th}$ spin  and vice-versa.

The zero quantum and double quantum elements are observed by using the following pulse sequence,
\begin{align}
A. ~&\bigg[\frac{\pi}{2}\bigg]_\alpha^i \bigg[\frac{\pi}{2}\bigg]_\beta^j - \bigg[G\bigg]_z - \bigg[\frac{\pi}{2}\bigg]_y^i, \\
B. ~&\bigg[\frac{\pi}{2}\bigg]_\alpha^i \bigg[\frac{\pi}{2}\bigg]_\beta^j - \bigg[G\bigg]_z -\bigg[\pi\bigg]^j \bigg[\frac{\pi}{2}\bigg]_y^i.
\end{align}
The experiments $A$ and $B$ are repeated four times with the phases $\alpha$, $\beta$ as $-y$, $-y$; $x$, $x$; $-y$, $x$; and $x$, $-y$. Here, the first two $\pi/2$ pulses convert the selected double quantum and zero quantum elements along with some additional single quantum elements into the diagonal part. Except these, all the other elements are dephased  by the followed gradient pulse. The final $\pi/2$ pulse converts these diagonal elements into the single quantum, which are observable. The additional single quantum elements which were picked up by the first two $\pi/2$ pulses can be filtered out by taking a linear combination of the spectra from the experiments $A$ and $B$, thus measuring only zero and double quantum elements.
\begin{figure} 
\hspace{-0.5cm}
 \centering
 \includegraphics[width=9cm]{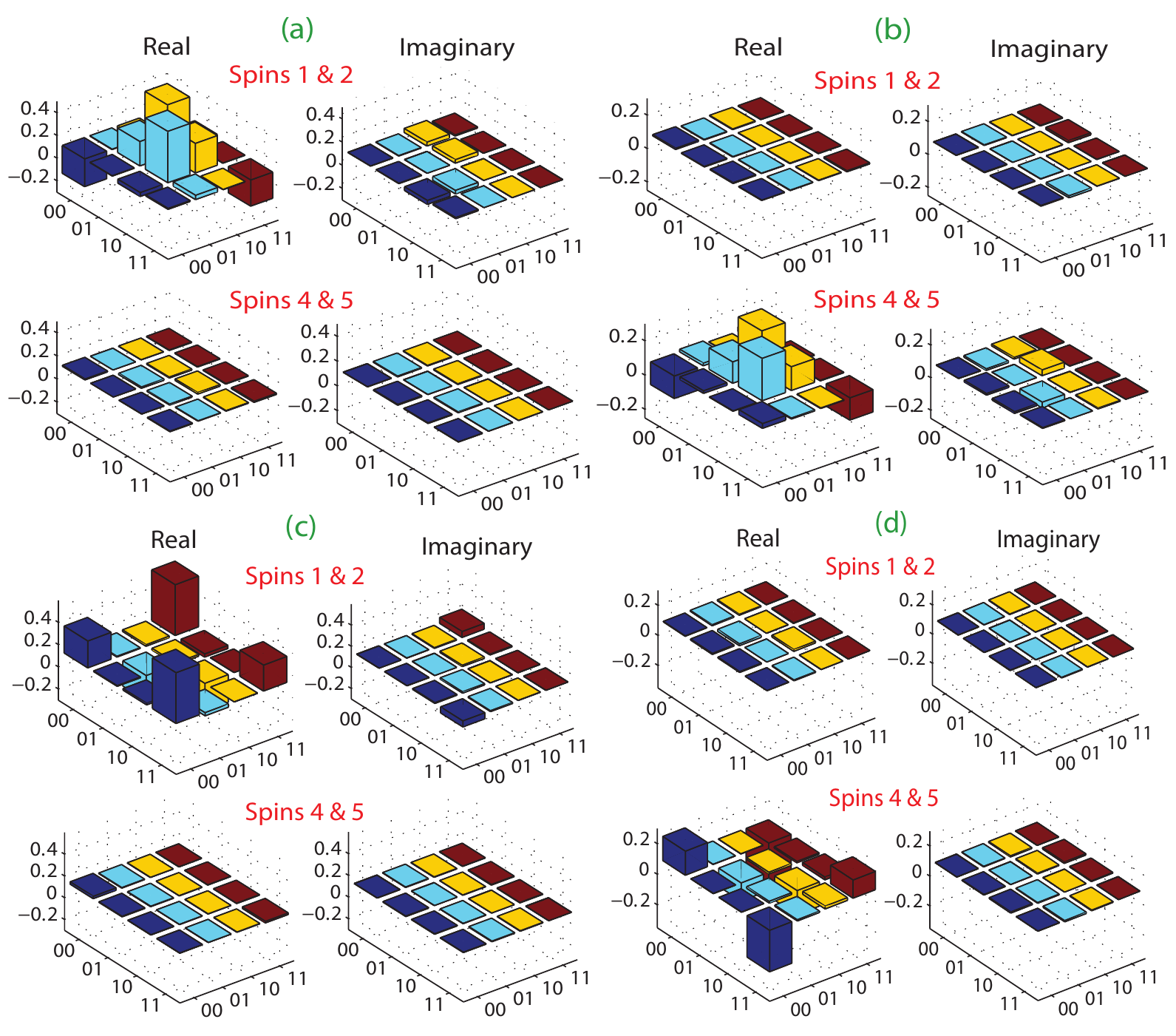}
 \caption{Experimental results for entanglement transfer. Reconstructed deviation density matrices (trace less) of spins $1$ and $2$, and spins $4$ and $5$ for (a) the initial state $\tfrac{1}{{2}} (\ket{01}+\ket{10})(\bra{01}+\bra{10})_{12} \otimes \mathbbm{1}_3 \otimes \mathbbm{1}_4 \otimes \mathbbm{1}_5$, (b) the final state $\mathbbm{1}_1 \otimes \mathbbm{1}_2 \otimes \mathbbm{1}_3 \otimes \tfrac{1}{{2}} (\ket{01}+\ket{10})(\bra{01}+\bra{10})_{45}$, (c) the initial state $\tfrac{1}{{2}} (\ket{00}+\ket{11})(\bra{00}+\bra{11})_{12} \otimes \mathbbm{1}_3 \otimes \mathbbm{1}_4 \otimes \mathbbm{1}_5$, and (d) the final state $\mathbbm{1}_1 \otimes \mathbbm{1}_2 \otimes \mathbbm{1}_3 \otimes \tfrac{1}{{2}} (\ket{00}-\ket{11})(\bra{00}-\bra{11})_{45}$.}
 \label{fig3} 
\end{figure}

The tomography results for both the initial and final states are shown in the Fig. (\ref{fig3}). 
We used two different measures to compare the experimental density matrices ($\rho_{\mathrm{expt}}$) with the theoretical density matrices ($\rho_{\mathrm{expt}}$), (i) fidelity ($F$), given by
\begin{equation}
F=\frac{\textrm{tr}(\rho_{\mathrm{th}} \rho_{\mathrm{expt}})}{\sqrt{\textrm{tr} (\rho_{\mathrm{th}}^2) \textrm{tr} (\rho_{\mathrm{expt}}^2)}},
\end{equation}
and (ii) attenuated correlation ($c$) \cite{Attcor}, given by
\begin{eqnarray}
c=\frac{\textrm{tr}(\rho_{\mathrm{th}} \rho_{\mathrm{expt}})}{\textrm{tr} (\rho_{\mathrm{th}}^2)}.
\end{eqnarray}
The attenuated correlation also accounts for the net loss of magnetization due to random errors along with the systematic errors.
%To quantitatively evaluate the experimental results, we calculate the fidelity (\(F\)) and attenuated correlation ($c$) of the experimental density matrices (\(\rho_{exp}\)) with 
%respect to the theoretical density matrices (\(\rho_{th}\)), given by
%\begin{eqnarray}
%F&=&\frac{\textrm{tr}(\rho_{th} \rho_{exp})}{\sqrt{\textrm{tr} (\rho_{th}^2) \textrm{tr} (\rho_{exp}^2)}},\\
%c&=&\frac{\textrm{tr}(\rho_{th} \rho_{exp})}{\textrm{tr} (\rho_{th}^2)}.
%\end{eqnarray}
\begin{table}
\centering
\includegraphics[width=8cm]{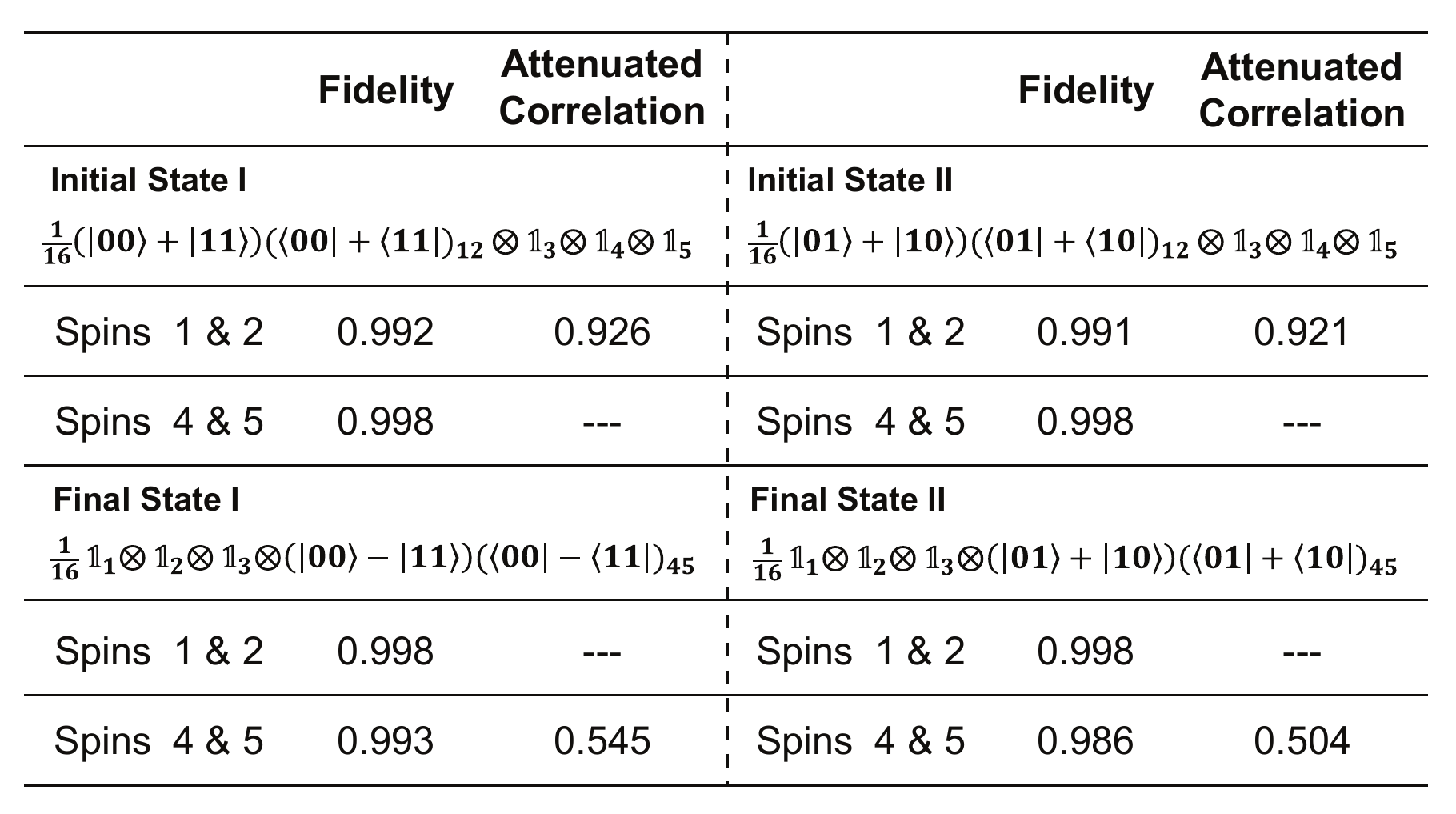}
\caption{Fidelity and attenuated correlation of all the initial and final entangled states.}
\label{tab1}
\end{table}
The fidelities and attenuated correlations of all the initial and final states are given in the table \ref{tab1}. The high fidelities of both the initial and final states indicate the efficient creation and transfer of entangled states. The low attenuated correlations of the final states are mainly due to the decoherence as the total time of the experiment ($\approx$ 52 ms including the preparation of initial state) is comparable to the $T_2^*$ of the fluorine spins.

\subsection{Four-spin pseudo-pure initial states}
\begin{figure}
 \centering
 \includegraphics[width=8cm]{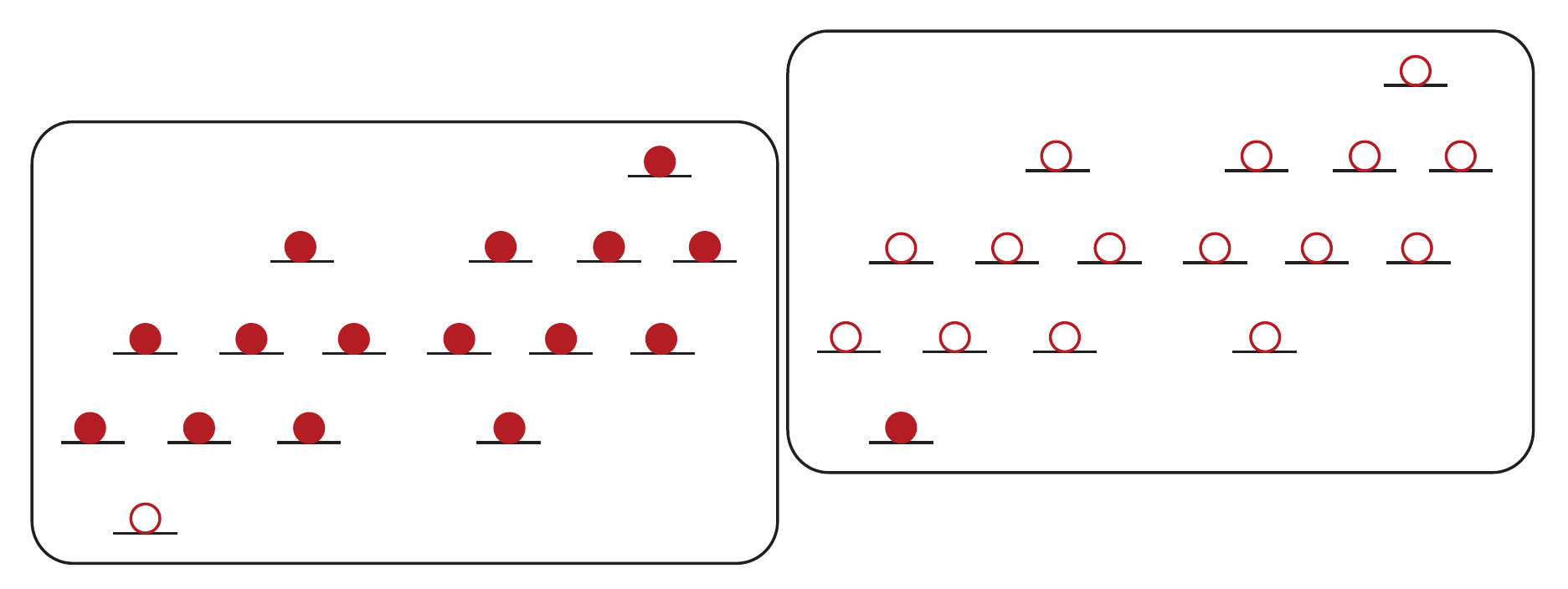}
 \caption{A representative energy level diagram and deviation populations for the $\ket{0000}$ pseudo-pure state. The filled and open circles represent the positive and negative deviation populations (w. r. t. the background population) respectively.}
 \label{figa} 
\end{figure}

%We now discuss the experimental implementation of mirror inversion operation on pseudo-pure initial states. 
Spatially Averaged Logical Labelling Technique (SALLT) \cite{SALLT} was used to prepare 4-spin pseudo-pure states in a 5-spin system as described below. 
Here, the Hilbert-space of the 5-spin system is divided into two 4-spin subspaces based on
the $\ket{0}$ and $\ket{1}$ states of spin $5$, which does not take part in the mirror-inversion operation.
%Here, spin $5$ is used for labelling the two subsystems formed by the other spins. 
Starting from equilibrium, we dephase the magnetization of all spins except spin $5$, by using the procedure described in the previous section. Then, the state of the system can be described by the density matrix $\sigma_5^z$. Now, the desired 4-spin pseudo-pure states (in each of the two subspaces corresponding to the $\ket{0}$ and $\ket{1}$ states of spin $5$) are prepared by flipping ($\pi$-rotation) the corresponding transition of spin $5$. For example, $\ket{0000}$ pseudo-pure state (labelled by the states of spins $1$ to $4$) is prepared by inverting the $\ket{00000} \leftrightarrow \ket{00001}$ transition of spin $5$.
A representative diagram for the deviation populations of the $\ket{0000}$ pseudo-pure state is given in the Fig. (\ref{figa}). Note that the deviation populations in both the subspaces (of spin $5$) are the same but are opposite in sign.

By using the above method, we prepared the $\ket{1000}\bra{1000}$, $\ket{1010}\bra{1010}$, and $\ket{1110}\bra{1110}$ initial pseudo-pure states. All the transition selective $\pi$ pulses were implemented by Gaussian shaped pulses of duration 40 ms. The above initial states were transformed into their mirror images $\ket{0001}\bra{0001}$, $\ket{0101}\bra{0101}$, and $\ket{0111}\bra{0111}$ respectively by the unitary operator in Eq. (\ref{Uop4}).

\begin{figure}
\hspace{-12.5pt}
 \centering
 \includegraphics[width=9cm]{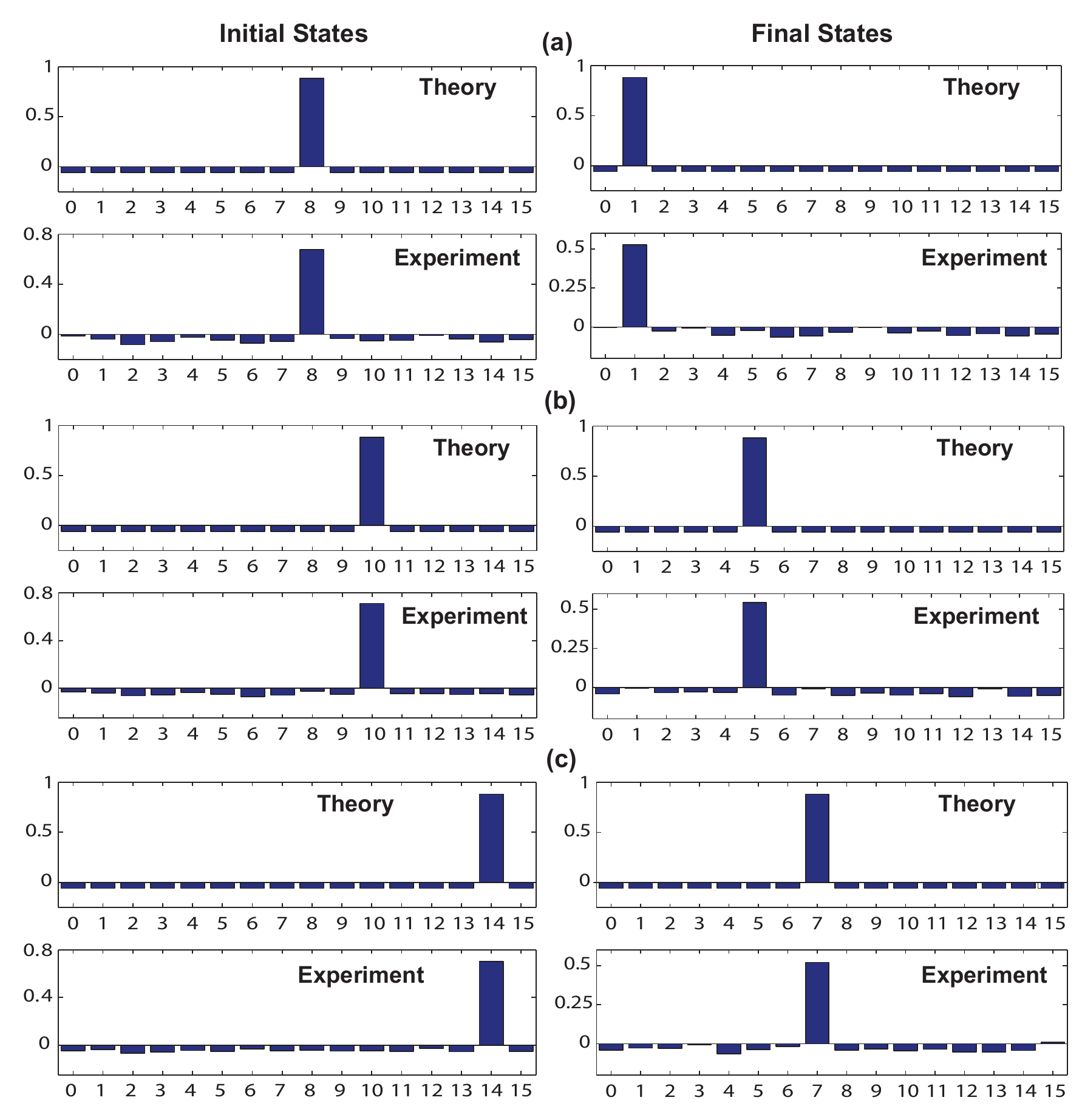}
 \caption{The theoretically expected and the experimentally reconstructed diagonal deviation density matrices (traceless; top and bottom rows in each case) for the initial (left column) and final (mirror-inverted; right column) pseudo-pure states, (a) $\ket{1000}\bra{1000} \rightarrow \ket{0001}\bra{0001}$, 
 (b) $\ket{1010}\bra{1010} \rightarrow \ket{0101}\bra{0101}$, and 
 (c) $\ket{1110}\bra{1110} \rightarrow \ket{0111}\bra{0111}$.}
% These bar plots are obtained by taking average over the deviation populations of the two subspaces of the spin $5$, where the sign of deviation populations of one of the two subspaces is reversed.}
% In all the cases, the top and bottom rows correspond to the theoretically expected and experimental results and the left and right columns correspond to the initial and final states respectively.}
 
% The reconstructed diagonal part of the deviation density matrices (traceless) along with the theoretically expected ones for (a) initial state $\ket{1000}\bra{1000}$ and final state $\ket{0001}\bra{0001}$, (b) initial state $\ket{1010}\bra{1010}$ and final state $\ket{0101}\bra{0101}$, (c) initial state $\ket{1110}\bra{1110}$ and final state $\ket{0111}\bra{0111}$. These are obtained by taking average over the two subspaces of the $5^{th}$ spin. In all the cases, the top row corresponds to the theoretically expected results and the left and right columns correspond to the initial and final states respectively.}
 \label{fig4} 
\end{figure}

Diagonal tomography of all the initial and final states were performed by applying a gradient pulse followed by a $\pi/2$ pulse on each spin separately and fitting the resultant single quantum spectra. The results along with the theoretically expected ones are shown  in Fig. (\ref{fig4}).
The bar plots shown in the figure represent the diagonal deviation density matrices (traceless) of spins $1$ to $4$.
These are obtained by taking average over the deviation populations of the two subspaces of spin $5$, where the sign of deviation populations of one of the two subspaces is reversed, and then subtracting the trace.
%The bar plots shown in the figure correspond to the results obtained by taking average over the deviation populations of the two subspaces of spin $5$. 
%While taking the average, the sign of the deviation populations of one of the two subspaces is reversed.
%represent the traceless diagonal density matrices of spins $1$ to $4$, which are obtained by taking average over the two subspaces of spin $5$.
The fidelities of the experimental diagonal density matrices with respect to the theoretically expected ones are calculated and the results are as follows. The diagonal fidelities of all the initial states are better than 0.994 and those of all the final states are better than 0.989.
%The slight decrease in the fidelities of these experiments compared to those of 5-spin initial states is mainly due to the errors in the implementation of transition selective $\pi$-rotations, used to prepare the pseudo-pure initial states.
%The low polarization of the final states compared to the initial states is mainly due to decoherence as again the total experimental time is comparable to the $T_2^*$ of fluorine spins.
The high fidelity of the final states confirms the successful implementation of mirror inversion operation on pseudo-pure initial states. 

\section{Conclusions}
\label{sec-conclu}

Mirror inversion of quantum states in spin chains has interesting applications in quantum information transport. Since the mirror inversion can be used to transfer states of multiple qubits including non-trivial entangled states, the transfer of data in quantum registers can be significantly simplified.
In this work, we performed experimental quantum simulation of mirror inversion operation in an XY spin chain with pre-engineered couplings. The experiments have been performed with a 5-qubit dipolar coupled spin system using NMR techniques. By using this simulation, we have demonstrated (i) transfer of quantum states of single qubits to their mirror images, and (ii) transfer of entangled states of multiple qubits to their mirror images. Due to the large dipolar couplings of the system, the multiqubit gates were constructed in lesser time and hence the experiments were performed more efficiently (with less experimental errors). The high fidelities of the present experiments imply that these experiments are possible for larger spin chains with existing control techniques if suitable systems are found.

The above quantum simulations are assisted by efficient decomposition of the unitary
evolution of XY spin-chain in the Pauli-operator basis.
%We have also discussed the decomposition of unitary operators into Pauli-operator basis.
The number of multiqubit operators 
%(for example, $e^{-i\frac{\pi}{4}\sigma_1^x \sigma_2^z \sigma_3^z \cdots \sigma_{N-1}^z 
%\sigma_N^x}$) 
in such a product decomposition
%$U_{XY}(\frac{\pi}{2})$ corresponding to the $N$-spin chain (Eqs. (\ref{Uoddn}), (\ref{Uevenn})) 
increases only linearly with the number of spins in the chain. 
Again, each of these multiqubit operators can be decomposed into 
local gates and two-qubit gates, whose numbers also scale linearly with the number of qubits.
%(for example, $e^{-i\frac{\pi}{4}\sigma_1^x \sigma_2^z \sigma_3^z \sigma_4^x}=e^{-i\frac{\pi}{4}\sigma_1^x \sigma_2^x} e^{-i\frac{\pi}{4}\sigma_3^x \sigma_4^x} e^{-i\frac{\pi}{4}\sigma_2^y \sigma_3^y} e^{i\frac{\pi}{4}\sigma_3^x \sigma_4^x} e^{i\frac{\pi}{4}\sigma_1^x \sigma_2^x}$).
%The number of two-qubit gates in such a decomposition also scales linearly with $N$. 
This implies that, in any experimental set-up, where the two-qubit gates
%, like $e^{-i\frac{\pi}{4}\sigma_1^x \sigma_2^x}$,
can be implemented efficiently, this mirror inversion operation can 
be simulated efficiently. In the case of NMR, if one uses the subsystems approach given in references \cite{Mahesh12, RyanPRA} along with the methods presented here, these simulations may be extended to larger spin systems. Overall, we believe, the methods presented here will be useful in quantum simulations by larger spin systems.

\section*{Acknowledgements}

We thank Abhishek Shukla for sample preparation, and Ashok Ajoy and Hemant Katiyar for helpful discussions. The use of AV500 FTNMR spectrometer of the NMR Research Center at IISER, Pune is gratefully acknowledged. This work was partly supported by DST projects IR/S2/PU-01/2008 and SR/S2/LOP-0017/2009.

\bibliography{bibMirInv}

\end{document}